\documentclass[aps,pra,twocolumn,showpacs,superscriptaddress]{revtex4}
\usepackage{longtable}
\usepackage{graphicx}
\usepackage{epsfig}
\usepackage{times}
\usepackage{amssymb}
\usepackage{color}

\newcommand{\beq}{\begin{equation}}
\newcommand{\eeq}{\end{equation}}
\newcommand{\bea}{\begin{eqnarray}}
\newcommand{\eea}{\end{eqnarray}}

\newcommand{\coon}{(Color online) }

\begin{document}

%\preprint{}

\title{Demonstration of a squeezed light enhanced power- and signal-recycled Michelson interferometer}

\author{Henning Vahlbruch}
\author{Simon Chelkowski}
\author{Boris Hage}
\author{Alexander Franzen}
\author{Karsten Danzmann}
\author{Roman Schnabel}
\affiliation{Institut f\"ur Atom- und Molek\"ulphysik, Universit\"at Hannover and Max-Planck-Institut f\"ur Gravitationsphysik (Albert-Einstein-Institut), Callinstr. 38, 30167 Hannover, Germany}

% repeat the \author .. \affiliation  etc.\ as needed

%\author{}
%\thanks{}
%\affiliation{}


\date{\today}

\begin{abstract}
We report on the experimental combination of three advanced interferometer techniques for gravitational wave detection, namely {\emph {power-recycling}},  {\emph {detuned signal-recycling}} and  {\emph {squeezed field injection}}. For the first time we experimentally prove the compatibility of especially the latter two.  To achieve a broadband non-classical sensitivity improvement we applied a filter cavity for compensation of quadrature rotation. Signal to noise ratio was improved by up to 
$2.8$~dB beyond the coherent state's shot noise. The complete set-up was stably locked for arbitrary times and characterized by injected single-sideband modulation fields.
\end{abstract}

\maketitle

%\section{Introduction}
Gravitational waves (GWs) have been predicted by Albert Einstein using the theory of general relativity, but so far they have not been observed \cite{Thorne87}. An international array of ground-based, kilometer-scale Michelson interferometers has been set up for their first observation, consisting of GEO\,600~\cite{geo02}, LIGO~\cite{LIGO}, TAMA\,300~\cite{TAMA} and VIRGO~\cite{VIRGO04}. The goal is to measure a gravitational wave induced strain of space-time of the order of $10^{-21}$ integrated over a bandwidth of a few hundred Hertz at acoustic frequencies.
Even for kilometer-scale interferometers the expected signals are that small that several kilowatts of circulating single mode laser radiation are required to push the shot noise below the signal strength. 
Such high powers cannot be achieved by today's lasers alone.
Power-recycling \cite{DHKHFMW83pr} is an advanced interferometer technique that aims for increased circulating power. Signal-recycling \cite{Mee88} was also invented to improve the signal-to-shot-noise ratio at some detection frequencies. In fact also the injection of squeezed states was first proposed to 
reduce shot-noise \cite{Cav81}.
Later in the 1980s it was realized that squeezed states can also be used to reduce the overall quantum noise in interferometers including radiation pressure noise, thereby beating the standard-quantum-limit
\cite{Unruh82,JRe90}.
Recently it was discovered that at radiation pressure dominated frequencies signal-recycling can also be used to beat the SQL \cite{BCh01a}.
Gea-Banacloche and Leuchs showed that the techniques of power-recycling and squeezed field injection are fully compatible \cite{GLe87}. 
Chickarmane {\it et al.}\,\cite{CDh96} found compatibility of signal recycling and squeezed field injection for the shot-noise limited regime.
Furthermore the analysis by Harms  {\it et al.} showed that the same is true for detuned signal-recycling at shot-noise as well as radiation pressure noise dominated frequencies \cite{HCCFVDS03} thereby proposing that all the three techniques can simultaneously been used to reduce quantum noise in interferometers.

All of the GW detectors mentioned above use power-recycling, additionally most of them also use arm cavities for further power build up. In both cases additional mirrors form \emph{tuned} cavities.
The GEO\,600 detector already successfully uses carrier light  \emph{detuned} signal-recycling. Signal recycling is established by an additional mirror placed into the interferometer's dark signal port forming a signal tuned cavity.This leads to an optical resonance structure in the interferometer's signal transfer function, whose frequency can be changed and matched to an expected signal, for example emitted by a binary system of two neutron stars or black holes. Signal-recycling in combination with power-recycling is often called dual-recycling and was experimentally demonstrated by Heinzel {\it et al.}\,\cite{HSMSWWSRD98}. 
Second-generation detectors currently being planned, for example Advanced LIGO \cite{ADVLIGO}, are likely to use this technique. Then in combination with tuned high finesse arm cavities the technique is called resonant sideband extraction \cite{MSNCSRWD93,HMSRWD96}.
Squeezed states are envisaged for third generation detectors but so far only a few squeezed light enhanced interferometers have been demonstrated, e.g.\,table-top Mach-Zehnder and polarization interferometers \cite{XWK87,GSYL87}, respectively. Recently a squeezing enhanced power-recycled Michelson interferometer has been reported already bearing more resemblance to a GW detector \cite{KSMBL02}. 

In this letter we report the first power- and signal-recycled Michelson interferometer with broadband sensitivity better than its photon shot-noise. 
Frequency dependent squeezed light generated in an optical parametric amplifier (OPA) in combination with a detuned filter cavity (FC), as proposed in \cite{KLMTV01}, was injected through the signal recycling mirror (SRM) into the interferometer«s dark port.
The whole setup was stably locked in all its degrees of freedom and characterized by a single-sideband modulation field. Our results generally prove the optical compatibility of the three advanced interferometer techniques described above and also demonstrates a readout and control scheme in which no phase modulation control signals contaminate the detection band.

%\section{Experimental}

\begin{figure*}[t!]
\hspace{-50mm}
\includegraphics[width=10cm]{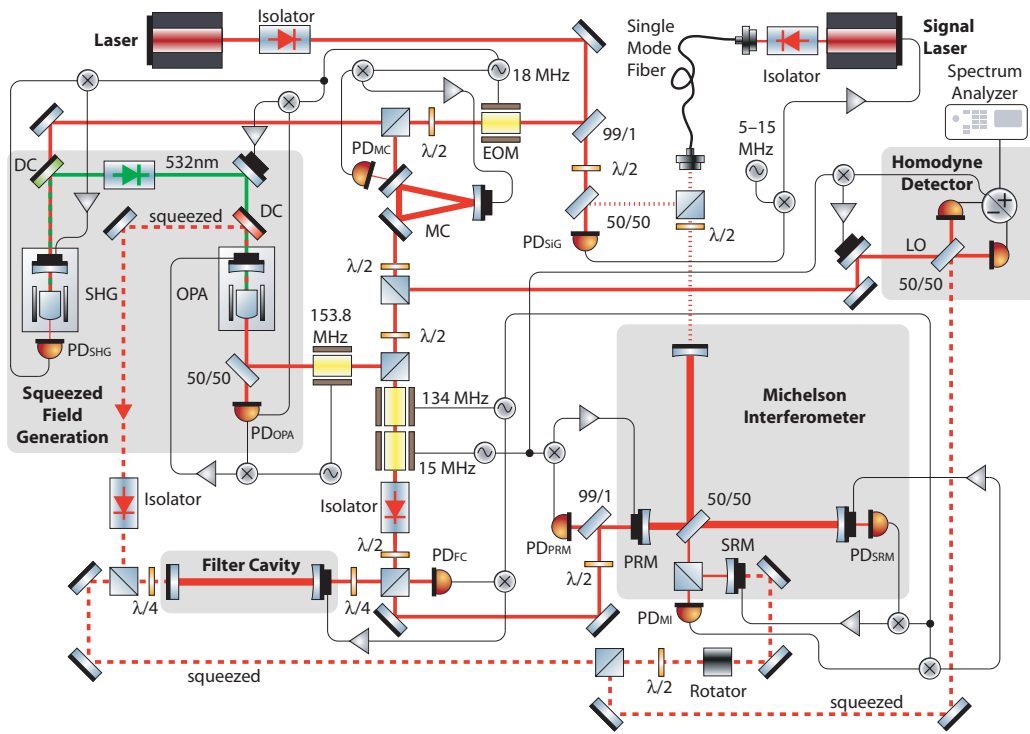}
  \vspace{0mm}
\caption{\coon Schematic of the experiment. Amplitude
squeezed light is generated in an OPA cavity of controlled length.
The detuned filter cavity provides frequency dependent squeezing
suitable for a broadband quantum noise reduction of a shot-noise
limited dual-recycled Michelson interferometer. SHG: second harmonic generation; OPA: optical
parametric amplifier; EOM: electro optical modulator; DC: dichroic mirror; LO: local oscillator; 
PD: photo diode; MC: mode cleaner; PRM: power recycling mirror; SRM: signal recycling mirror;
\vrule width 3mm depth -.6mm\,: piezo-electric transducer.}
  \label{experiment}
\end{figure*}

The main laser source of our experiment was a monolithic non-planar Nd:YAG ring laser of 2\,W single mode output power at 1064 nm. About 1\,W was used for second harmonic generation (SHG) to produce the necessary pump field for the optical parametric amplifier (OPA), see Fig.~\ref{experiment}. 
The residual beam was transmitted through a mode cleaner ring cavity to reduce laser amplitude noise and spatial fluctuations. The outgoing field was used as a local oscillator for the homodyne
detector (50\,mW) as well as a seed beam for the OPA (100\,mW), filter cavity locking  (30\,mW) and the Michelson (120\,mW). For controlling the OPA cavity length a Pound-Drever-Hall (PDH) locking scheme with a phase modulation sideband frequency of 153.8\,MHz was used.The error signal was fed back to the PZT mounted coupling mirror of the hemilithic cavity. A similar locking technique was used for the SHG. Another locking loop stabilized the phase relation between the
fundamental and second harmonic field inside the OPA. A more detailed description can be found in \cite{CVHFLDS05}. Locking the OPA to deamplification generates a broadband amplitude quadrature squeezed beam of about 200\,$\mu$W at 1064\,nm.
This beam was then first passed through a Faraday isolator, protecting the OPA from any backscattered
light. A $\lambda/4$-waveplate turned the s-polarized beam into a circularly polarized beam which was mode matched into our linear filter cavity (FC).
The coupling and end mirrors had reflectivities of 90\,\%%
\,\,and 99.92\,\%, respectively.
The cavity length was electronically stabilized to about L=1.21\,m resulting in a free spectral range of
124\,MHz. We applied the PDH locking technique utilizing a circularly polarized laser beam
that carried 134\,MHz phase modulation sidebands and was coupled into the filter cavity from the opposite side. Therefore it was possible to lock the FC stably to a sideband frequency of
$\pm$134\,MHz which results  in a detuning frequency of $\pm$10\,MHz due to the free spectral range of 124\,MHz. This technique avoided unwanted control signals showing up in the detection band. Note that such a signal was present in Figs.\,3a and 4a of Ref.\,\cite{CVHFLDS05} at 15\,MHz. 
Locking the filter cavity to either the upper or the lower sideband the squeezed field was then
reflected towards the signal recycling mirror of the
Michelson interferometer.

The Michelson interferometer was dual recycled; both recycling cavities had lengths of about 1.21\,m and the reflectivities of power recycling mirror (PRM) and signal recycling mirror (SRM) were both 90\,\%, cf. Fig.\,\ref{experiment}. The interferometer was stabilized on a dark fringe and the PRM was controlled such that it formed a carrier field resonating cavity together with the two Michelson end mirrors of 99.92\,\% reflectivity. The finesse of this power recycling cavity (PRC) was measured to 60. 
The signal recycling cavity that was formed by SRM and the two end mirrors contained no carrier field and could be stably locked to sideband frequencies of $\pm$10\,MHz. 
All together three electronic control loops were applied to stabilize the interferometer to this operation point utilizing two polarization modes and two modulation frequencies. 
The phase modulations at 15\,MHz and 134\,MHz were applied to the carrier
field before it entered the interferometer and a $\lambda/2$-plate
in front of the PRM split the incoming beam into 100\,mW of
s-polarization and 20\,mW of p-polarization. The position of the PRM was
locked by a PDH technique via the 15\,MHz sidebands in the s-polarization.
Modulation sidebands at 134\,MHz in the s- and p-polarizations were then used to control the dark port and the length of the detuned signal recycling cavity, respectively. The polarization modes were decoupled by a polarizing beam splitter (PBS)  that was placed between the 50/50 beamsplitter and the SRM. An arm length difference of the Michelson of 7\,mm was sufficient to provide adequate strong error signals.

\begin{figure}[t!]
\hspace{0mm}
\includegraphics[angle=0,width=7cm]{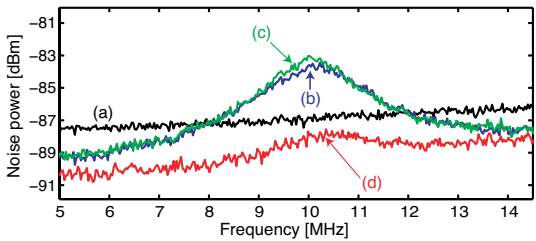}
\caption{\coon Amplitude quadrature noise power spectra from homodyne detection: (a) shot noise,  (b) frequency dependent squeezing after reflection at the $+$10\,MHz detuned signal-recycling cavity (SRC) only, (c) frequency dependent squeezing after reflection at the $-$10\,MHz detuned filter cavity (FC) only and (d) regained broadband squeezed light enhanced performance achieved by reflecting the squeezed light firstly at the FC before it entered the SRC.}
 \label{Filtereffekt}
\end{figure}

The squeezed beam from the detuned locked filter cavity was injected into the SRC passing a combination of a PBS, a $\lambda/2$-plate and a Faraday rotator. This gave spatial degeneracy between the reflected squeezing and the signal output beam of the interferometer.
The combined field was guided to a homodyne detector that was built from two electronically and optically matched  photodetectors based on Epitaxx ETX1000 photodiodes. All spectra presented in this paper were analyzed in a Rohde\&Schwarz FSP3 spectrum analyzer with 100\,kHz  resolution bandwidth and 100\,Hz video bandwidth, averaging over 5 subsequent measurements.

The setup described so far facilitate the measurement of the interferometer's noise transfer function. The measurement of the signal transfer function and therefore the combined signal-to-noise ratio required the generation of a modulation signal of known strength and tunable sideband frequency. We chose to inject a single sideband modulation field into the interferometer, similar in design to \cite{GdeVine}. Such a field is generically different from a phase modulation, for example generated by a gravitational wave. 
However, it has been shown in \cite{SHSD04} that a shot-noise limited signal-recycled interferometer with detuning larger than the SRC bandwidth cannot be significantly improved by frequency dependent homodyning, so-called variational output. The reason is that only a single sideband is supported and injecting a single sideband modulation is therefore a meaningful method to characterize the signal-recycled interferometer.
The signal was generated utilizing a second monolithic non-planar Nd:YAG ring
laser. This laser was frequency locked to the main laser by a phase lock loop,
with tunable beat frequency in the range of 5--15\,MHz between both light sources. This signal laser beam was injected through one of the interferometer end mirrors as shown in Fig.~\ref{experiment}.

\begin{figure}[t!]
\hspace{0mm}
\includegraphics[angle=0, width=7cm]{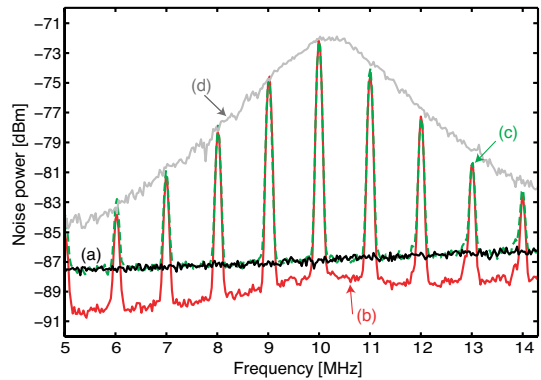}
\caption{ \coon Amplitude quadrature power spectra of the dual recycled Michelson interferometer with and without nonclassical noise reduction.
A single sideband modulation was injected at certain frequencies to characterize the signal to noise ratio. (a) Shot-noise measured with a blocked signal beam at the homodyne detector; (b) (dashed line) shot noise limited signals; (c) broadband squeezing enhanced signal-to-noise ratios of up to 2.8\,dB below shot-noise.  Utilizing the max hold function of the spectrum analyzer the optical transfer function of the SRC was mapped out by continuously sweeping the injected signal (d).}
\label{SNR}
\end{figure}

%\section{results and discussion}

Fig.~\ref{Filtereffekt} shows the squeezed light enhanced quantum noise performance of our dual-recycled interferometer. By using frequency dependent squeezed light the quantum noise can be reduced below the shot noise level (a) for all sideband frequencies shown (d). If instead no filter cavity is used the quantum noise is increased at some sideband frequencies (c). This effect is a consequence of a rotation of the field's quadratures when reflected from a detuned cavity \cite{KLMTV01}. While the OPA generated an amplitude squeezed beam the squeezing ellipse is rotated such that antisqueezing is detected around the SRC-detuning frequency by the homodyne detector.
This effect has first been observed and characterized in \cite{CVHFLDS05}.
The filter cavity compensates this effect while being locked to the same detuning frequency but with opposite sign. The remaining hump around 10\,MHz in curve (d) arises from additional losses inside the SRC mainly caused by the PBS. For this reason the squeezing got reduced most exactly at the detuning frequency.

The mode matchings of the squeezed field into the FC and SRC were 95\,\% and 97\,\%, respectively. Additional losses on the squeezing arose from the OPA escape efficiency of 90\,\%, transmittance of the isolator including double passing the Faraday rotator of 93\,\%, 
modematching efficiency at the homodyne detector of 95\,\% and 
quantum efficiency of photo diodes of 93\,\% adding up to an overall efficiency of 65\,\%. 
The result was a detected nonclassical noise supression of 2.8\,dB at 5\,MHz.
The poorer squeezing of about 2.0\,dB at 14\,MHz was due to the limited bandwidth of the OPA of 20\,MHz. In principle the bandwith can be increased by higher coupler transmission which then in turn requires a higher pump intensity, but since in GW detectors squeezing at acoustic frequencies will apply the OPA bandwidth limitation will not be an issue. We note that all measured spectra were at least 5\,dB above the detection dark noise which was taken into account.

Fig.~\ref{SNR} demonstrates the squeezing enhanced signal to noise ratio of single sideband signals. 
The Michelson interferometer was stably locked with $+10$\,MHz detuned signal-recycling cavity, while a single sideband was subsequently injected at ten different sideband frequencies and the spectra were measured with and without frequency dependent squeezed field injection. The experimental results show an increase of the SNR over the whole detection bandwidth.
In comparison with the shot noise limited signals (c) the SNR was improved by up to 2.8\,dB  into the nonclassical regime (b). 

%\section{conclusion}

In conclusion we have experimentally demonstrated the compatibility of power-recycling, carrier detuned signal-recycling, and squeezed field injection. Broadband nonclassical noise suppression was achieved by employing a detuned filter cavity for compensation of quadrature rotation. The optical layout of our demonstration experiment can directly be applied to improve the sensitivity of large scale signal-recycled interferometers at their shot-noise limited detection frequencies, typically above 1\,kHz. 
The scheme demonstrated also directly applies to the shot-noise limited resonant-sideband extraction topology which is planned for the Advanced LIGO detector (LIGO\,II) \cite{ADVLIGO}. Then the filter cavity needs to be adapted to compensate for the quadrature rotation due to the RSE cavity \cite{HCCFVDS03}.

After applying classical noise suppression to enable squeezing in the 
GW band as demonstrated in \cite{MGBWGML04}  our squeezed light source would be directly applicable to all current and probably next generation detectors since the same laser 
wavelength is used.
Future detectors are also expected to be quantum noise limited at lower frequencies due to back-action noise (radiation-pressure-noise). Then the demonstrated experiment can also provide a broadband nonclassical sensitivity improvement if additional filter cavities are used. Then if the initial power- and signal-recycled interferometer is already at the standard-quantum-limit, squeezed field injection can provide QND performance \cite{HCCFVDS03,KLMTV01,BC}. In our experiment we chose a homodyne readout with external local oscillator. This is in accordance with theoretical investigations that considered a readout scheme that can switch between arbitrary quadrature angles \cite{BCh01a,HCCFVDS03}. Such a local oscillator might be generated in a large scale GW detector from an unused reflection from the beam splitter anti-reflection coating. We believe that a realistic goal for future GW detectors is a 6\,dB improvement of power noise spectral density. This requires the generation of 10\,dB squeezed field at detection frequencies and an overall loss of less than 17\,\%. A detailed estimation of expected individual loss contributions in squeezing enhanced GW detectors will be presented elsewhere \cite{VCHFDS05b}.
However, if loss contributions due to the signal-recycling cavity or the filter cavity lead to an overall loss higher than that, only the performance at the optical resonance is degraded from the 6\,dB goal. The performance at neighbouring detection frequencies is then unaffected providing an improved detection bandwidth.

 This work has been supported by the Deutsche Forschungsgemeinschaft and is part of Sonderforschungsbereich 407.

\appendix

\end{document}